# CYCLOOXYGENASE INHIBITION IN ISCHEMIC BRAIN INJURY

## Eduardo Candelario-Jalil [1,*] and Bernd L. Fiebich [2]

[1]*Department of Neurology, University of New Mexico Health Sciences Center, Albuquerque, NM 87131, USA*
[2]*Neurochemistry Research Group, Department of Psychiatry, University of Freiburg Medical School, Hauptstrasse 5, D-79104 Freiburg, Germany*

*\* Author to whom all correspondence should be addressed:*
Eduardo Candelario-Jalil, PhD
Department of Neurology,
University of New Mexico Health Sciences Center,
MSC10 5620, Albuquerque, NM 87131-0001
USA
Tel: +1-(505)-925-4042
Fax: +1-(505)-272-6692
Email: ecandelario-jalil@salud.unm.edu

**Acknowledgments:** Authors are grateful to the Alexander von Humboldt Foundation (Bonn, Germany) and to the American Heart Association (Dallas, TX, USA) for supporting our studies.

## ABSTRACT
Neuroinflammation is one of the key pathological events involved in the progression of brain damage caused by cerebral ischemia. Metabolism of arachidonic acid through cyclooxygenase (COX) enzymes is known to be actively involved in the neuroinflammatory events leading to neuronal death after ischemia. Two isoforms of COX, termed COX-1 and COX-2, have been identified. Unlike COX-1, COX-2 expression is dramatically induced by ischemia and appears to be an effector of tissue damage. This review article will focus specifically on the involvement of COX isozymes in brain ischemia. We will discuss issues related to the biochemistry and selective pharmacological inhibition of COX enzymes, and further refer to their expression in the brain under normal conditions and following excitotoxicity and ischemic cerebral injury. We will review present knowledge of the relative contribution of each COX isoform to the brain ischemic pathology, based on data from investigations utilizing selective COX-1/COX-2 inhibitors and genetic knockout mouse models. The mechanisms of neurotoxicity associated with increased COX activity after ischemia will also be examined. Finally, we will provide a critical evaluation of the therapeutic potential of COX inhibitors in cerebral ischemia and discuss new targets downstream of COX with potential neuroprotective ability.

## INTRODUCTION

Cerebral ischemia is a leading cause of death and disability [1-4]. According to the World Health Organization (WHO), stroke is the second highest cause of mortality in the world and the third highest in developing countries. It is also a major cause of long-term disability worldwide [1;3;5-7], representing an important economic burden for society [8;9]. Very few therapeutic options are currently available [10], and there is an unmet need of neuroprotective agents to curtail the devastating cascade of pathological events triggered by the occlusion of a major brain vessel. Significant efforts have been devoted for decades to gain a better understanding of the pathophysiological mechanisms leading to neuronal death following ischemia. The search for neuroprotective compounds and novel techniques for timely injury detection, such as magnetic resonance imaging (MRI), positron emission tomography (PET), and computed tomography (CT) scans, is actively progressing and has led to significant advances in our understanding of stroke pathogenesis.

Among the damaging events following brain ischemia, excitotoxicity and alteration of intraneuronal $Ca^{2+}$ homeostasis are considered early processes that propagate the cascade of harmful processes leading to brain infarction [11-13]. Data published over the past decade have provided insights into the evolution of injury, and have suggested a greater role for oxidative stress and inflammatory mediators [14-16].

A wide body of evidence has recently accumulated supporting the notion that neuroinflammation plays a pivotal role in the progression of ischemic brain injury and pharmacological strategies aimed at reducing the inflammatory cascade following cerebral ischemia have proven to exert potent neuroprotection. Due to the concerted action of phospholipases, huge amounts of free arachidonic acid are released from membrane phospholipids after the ischemic event. The arachidonic acid metabolic pathway via cyclooxygenases is actively involved in neuroinflammation following the occlusion of cerebral blood vessels, and is a key component of the initiation and propagation of the inflammatory process after different types of brain injury including ischemia [17;18].

This review will focus specifically on the role of cyclooxygenase (COX) in ischemic brain injury. We first discuss aspects related to the biochemistry and pharmacological inhibition of COX, and further refer to the expression and role of each COX isozyme in excitotoxicity and ischemic cerebral injury. The mechanisms of toxicity of increased COX activity after brain ischemia are also examined. Finally, we provide a critical appraisal of the therapeutic potential of COX inhibitors in cerebral ischemia and discuss new potential targets downstream of COX with promising neuroprotective ability to ameliorate ischemic brain injury.

## BIOCHEMICAL ASPECTS OF CYCLOOXYGENASE

Cyclooxygenase, more correctly termed prostaglandin-endoperoxide synthase, catalyzes the conversion of arachidonic acid (AA) into prostaglandin $H_2$ ($PGH_2$) via $PGG_2$, a very short-lived endoperoxide intermediary [19-23]. $PGH_2$ is the precursor of several other prostanoids, including prostaglandin $E_2$ ($PGE_2$), prostacyclin ($PGI_2$), and thromboxanes (**Figure 1**). Although the more correct biochemical terminology for this enzyme is prostaglandin-endoperoxide synthase or prostaglandin G/H synthase (E.C. 1.14.99.1), its trivial name cyclooxygenase (COX) will be used throughout the present review article.

Biochemical and structural studies of COX have identified two different catalytic sites: the cyclooxygenase and the peroxidase reactions [19;24]. Free AA is cyclized and *bis*-oxygenated to yield $PGG_2$, an intermediary endoperoxide, in the cyclooxygenase site of the enzyme [19;24;25]. In a separate heme-containing peroxidase active site, a hydroperoxyl group in $PGG_2$ is reduced to hydroxyl forming $PGH_2$ [19;22]. It has been shown that free radical species are formed during the peroxidase reaction of COX (**Figure 1**), however, there is still debate on the identity of this free radical and the biological significance of its formation [22;23]. COX catalyzes the obliged reaction in the oxidative conversion of AA into different prostaglandins and thromboxanes, which are important in many physiological and pathological processes. It is important to mention that the production of $PGH_2$ is self-limited due to suicide inactivation of COX [26-28]. The biological significance of this biochemical process is presently unknown.

Once $PGH_2$ is formed, several tissue-specific isomerases are responsible for the synthesis of different biologically active prostanoids (**Figure 1**) [29-34]. For example, prostaglandin E synthase (PGES) metabolizes $PGH_2$ into $PGE_2$ [29;31]. There are several other terminal synthases including prostacyclin synthase (PGIS), prostaglandin $D_2$ synthase (PGD synthase), prostaglandin $F_{2\alpha}$ synthase and thromboxane synthase ($TXA_2$ synthase) [34]. This biochemical pathway is further complicated by the occurrence of several isoforms of some of the terminal synthases, and their differential expression in cell types during physiological and pathological conditions.

The biological effects of each prostanoid depend on its binding to receptors coupled to specific signal transduction mechanisms modulating cell function (**Figure 1**). For example, four different receptors for $PGE_2$ have been identified, EP1 through EP4 [35-37]. Interestingly, these EP receptors mediate different and sometimes opposing responses: EP2 and EP4 lead to increase cAMP [38], while EP3 activation reduces cAMP [39], and EP1 increases intracellular $Ca^{2+}$ [40;41]. Binding of $PGD_2$ to its DP1 receptor leads to increased cAMP levels [42;43]. These few examples serve to exemplify the complexity of prostanoid signaling, an area of research that investigators have recently started to discover. This information will be needed to better understand both the physiological and pathological effects of prostanoids.

Two different isoforms of COX, termed COX-1 and COX-2, have been cloned and extensively characterized [19;26;33;44]. COX-3, a putative isozyme of COX, was recently cloned, but it was found to be a splice variant of the transcript for COX-1 [45-47]. Great similarities between COX-1 and COX-2 have been observed [19]. They have an amino acid identity of approximately 60%, producing a 70 kDa protein that gets assembled within the lumen of the endoplasmic reticulum and nuclear envelope [19;33;44;48]. In terms of enzyme kinetics, both isoforms show very similar behavior [19;33;49]. Perhaps the most striking difference between COX-1 and COX-2 is their expression pattern. Data from different laboratories over several years have demonstrated the inducible nature of COX-2 as opposed to COX-1 [19;26;50-55]. The expression of COX-2 is upregulated by bacterial lipopolysaccharide (LPS), pro-inflammatory cytokines, growth factors, and tumor promoters, thereby distinguishing it from the primarily constitutive COX-1 isoform [52;54;56-62]. Due to this distinctive expression profile, COX-2 has been speculated to be the major isoform involved in inflammation and other pathologies.

The slightly enlarged COX-2 active site relative to COX-1 is essential in the recognition of bulkier substrates by COX-2 [44]. For example, COX-2, but not COX-1, effectively oxidizes the endocannabinoids anandamide (arachidonyl ethanolamide) and 2-arachidonyl glycerol, yielding prostanoid-like oxidized endocannabinoids [63-65]. The biological relevance of the COX-2-mediated metabolism of endocannabinoids is presently poorly understood.

Outstanding reviews of COX structure and kinetics, prostanoid receptors and terminal synthases have recently been written [19;29;31;33;35;44;66;67], and thus, in this section we have only discussed the essential aspects of these topics needed to understand the role of COX in cerebral ischemia.

## PHARMACOLOGICAL INHIBITION OF CYCLOOXYGENASE

Non-steroidal anti-inflammatory drugs (NSAIDs) are among the most used drugs worldwide. In early 1971, Sir John R. Vane proposed that NSAIDs inhibit prostaglandin synthesis, and thus act as anti-inflammatory and ulcerogenic agents [68;69]. It has now become widely recognized that nearly all NSAIDs (e.g., aspirin, ibuprofen, indomethacin, naproxen) act by blocking the access of the AA substrate to the COX active site [70;71]. With the major exception of aspirin, which acetylates and permanently inhibits COX, the other NSAIDs inhibit COX by competing with AA for binding in the COX active site [70;71]. Nevertheless, there are significant differences among NSAIDs in whether they bind the COX active site in a time-dependent or independent way.

Despite the similarities between COX-1 and COX-2 in terms of catalytic mechanisms and kinetics, there are two major structural differences that have important consequences from the pharmacological and biological viewpoint. First, COX-2 has a larger and more accommodating cyclooxygenase active site as compared with COX-1 [19;48]. This structural difference in the AA pocket size and shape has allowed the development of highly selective COX-2 inhibitors. Second, COX-1, as opposed to COX-2, exhibits negative allosterism at low AA concentrations. Thus, when both isoforms are expressed in the same cell, COX-2 competes more effectively for newly released AA [19].

The subtle structural differences between COX-1 and COX-2 have been exploited in order to develop selective inhibitors of each isoform [72;73]. This is a powerful pharmacological tool that, along with genetic knockout studies, has allowed the study of the specific physiological and pathological role of each COX isozyme.

There are excellent review articles of the cyclooxygenase catalysis and inhibition by NSAIDs [19;44;66;74-76], which the reader is encouraged to reference the very detailed, and at times complex, mechanisms of inhibition and pharmacology of highly selective inhibitors of each COX isoform.

## BASAL CYCLOOXYGENASE EXPRESSION IN THE BRAIN

Under normal conditions, COX-1 and COX-2 are expressed by specific neuronal populations [50]. COX-1 is widely expressed in neurons and microglia in several animal species [51;77-82], and its immunoreactivity is enriched in midbrain, pons and medulla [83]. COX-1 is intensely expressed in microglia in both rat [77] and human [51;84;85], and relative high levels are also found in hippocampal and cortical neurons in control human brain [51].

The brain is one of the few tissues that constitutively express COX-2 [86]. The immunoreactivity for COX-2 is primarily localized to neurons, and in the rat brain it prevails in the hippocampus with the densest staining in the dentate gyrus, piriform cortex, amygdala, and in layers II and III of the neocortex [81;83;87-89]. COX-2 astrocytic immunoreactivity has been described in the white matter of the rat brain and spinal cord [90;91].

Earlier studies have shown that synaptic stimuli induce neuronal COX-2 expression [87]. COX-2 expression colocalizes selectively with glutamatergic neurons [87;92], and seems to be coupled to excitatory neuronal activity [87;90;93-95]. It has been demonstrated that activation of the glutamatergic NMDA receptors results in increased neuronal COX-2 expression. Treatment with MK-801, a specific NMDA receptor antagonist, completely inhibited COX-2 induction following seizures [87].

Only limited information is available about the normal function of COX-2 in the brain. Due to its localization in dendrites and dendritic spines, and its upregulation following synaptic activation, it has been suggested that COX-2 plays a role in synaptic signaling and memory consolidation [17;50;87;92;96]. In the hippocampal formation, COX-2 also has been implicated in learning and memory [87;96-102]. Furthermore, COX-2 plays an important role in mediating the changes in neocortical blood flow evoked by neural activity [103]. There is a modulatory role of arachidonic acid metabolites produced by COX-2 in the control of focal cerebral blood flow following local increases in neuronal activity, suggesting the involvement of COX-2 in cerebrovascular coupling [104]. This is a very important homeostatic mechanism matching the delivery of nutrients with the energy needs of the active brain [105-107].

## EXPRESSION OF CYCLOOXYGENASES FOLLOWING EXCITOTOXICITY OR CEREBRAL ISCHEMIC INJURY

COX-2 is induced as an immediate early gene in a wide variety of cell types and in response to an ample variety of stimuli [44;50]. Since COX-2 expression is linked to synaptic activity, it is not surprising that its expression is dramatically increased following excitotoxic or ischemic injury.

### COX-2 expression following excitotoxicity

Dramatic increases in COX-2 mRNA and protein levels occur following *in vivo* excitotoxic injury [93;94;108-115]. Increased upregulation of COX-2 mRNA and protein has also been found in neuronal cultures exposed to excitotoxic insults *in vitro* [116;117]. On the contrary, COX-1 expression levels are not changed by excitotoxic damage [93;113]. The increased COX-2 expression following kainate excitotoxicity can be prevented by blockade of glutamate receptors [93], and by an antagonist of platelet-activating factor (PAF) receptor [95]. Substantial COX-2 overexpression was found in vulnerable neurons following kainic acid injection [94;111], and another study demonstrated localization of COX-2 in neurons undergoing apoptosis, as assessed using immunohistochemistry and TUNEL assay 8 h after kainic acid injection [110]. These reports provide evidence suggesting that COX-2 activity is closely linked to excitotoxic neuronal death. However, not all cells overexpressing COX-2 are destined to die following kainic acid injection. For example, a dramatic and prolonged increase in COX-2 immunoreactivity is seen in the dentate gyrus granule neurons, but these cells are not vulnerable to kainic acid neurotoxicity [111].

**COX-2 expression after brain ischemia**

It has long been known that COX-2 expression is significantly and persistently increased following cerebral ischemia in different models. In this section, we will describe the most important observations of the effect of brain ischemia on COX-2 expression in both global and focal cerebral ischemia.

*Expression of COX-2 following global cerebral ischemia*

Global cerebral ischemia results from the transient cessation of blood flow to the brain, leading to a characteristic pattern of cell death in specific neuronal populations. The hippocampal CA1 region is considered the most susceptible to a global ischemic event [118-120]. Clinically, global brain ischemia occurs after cardiopulmonary bypass surgery or cardiac arrest with resuscitation, and is associated with problems with cognition and memory, sensorimotor deficits, seizures, and death [121].

In a pioneer study, Ohtsuki and colleagues demonstrated induction of COX-2 mRNA in the gerbil hippocampus after transient global ischemia using Northern blotting and *in situ* hybridization [122]. Expression of COX-2 mRNA was markedly induced at 3 h of reperfusion, but after 24 h, the signal became undetectable. *In situ* hybridization revealed that both the pyramidal cells of the CA1-4 regions and the granular cells of the dentate gyrus expressed COX-2 up to 16 h after ischemia [122]. Occlusion of both common carotid arteries in gerbils caused delayed neuronal death in CA1 and CA2 areas 7 days later, but CA3 and dentate granular cells were preserved, indicating that COX-2 expression did not correlate with the resultant outcome of neuronal death after the ischemic insult in all areas. Additionally, COX-2 mRNA was also induced in many cortical neurons that were resistant to 5-min global cerebral ischemia [122].

In another study, both COX-2 mRNA and protein were substantially increased after ischemia in CA1 hippocampal neurons before their death in a rat model of 20 min of global cerebral ischemia. Twenty-four hours after ischemia, there was a large increase in hippocampal PGE$_2$ [123]. COX-2 protein induction peaked at 12-24 h after ischemia in several hippocampal regions, but in the CA1 neurons COX-2 was still seen at 3 days following global ischemia [124]. Also in a global ischemia model, bilateral occlusion of both common carotid arteries in gerbils resulted in a biphasic and significant increase in hippocampal PGE$_2$ concentrations (2 and 24-48 h of recirculation). The late increase in PGE$_2$ production preceded the onset of morphological changes in the CA1 subfield of the hippocampus [125]. Other studies have also confirmed previous observations of the significant increase in COX-2 expression following global ischemia [126-131].

*COX-2 expression in the brain subjected to focal cerebral ischemia*

Focal cerebral ischemia is produced when a major brain artery is blocked by a clot or bursts (brain hemorrhage). In the clinic, this type of injury is commonly referred to as ischemic or hemorrhagic stroke. In the industrialized countries, stroke is a leading cause of death and adult disability, representing a major public health problem [132].

A plethora of reports have shown that COX-2 expression is markedly upregulated following focal ischemia. Using a rat model of transient ischemic stroke induced by 1-h middle cerebral artery occlusion (MCAO) followed by reperfusion, Planas and co-workers demonstrated for the first time that COX-2 mRNA and protein are significantly induced in the ipsilateral cerebral cortex at 4 and 8 h after reperfusion, respectively [133]. These investigators also described colocalization of COX-2 and c-Fos immunoreactivities in superficial cortical layers within the ipsilateral hemisphere at 6 and 24 h following focal ischemia [111]. In permanent MCAO, a significant increase in COX-2 mRNA was detected in the core and penumbra regions of the cerebral cortex between 4 and 24 h after the occlusion, and this was most marked at 4 h in the penumbral areas [134]. In this study, a correlation was shown between the extent of COX-2 mRNA induction in the cortex at 4 h and the severity of brain damage detected at 24 h post MCAO [134]. Administration of MK-801, an NMDA antagonist, significantly attenuated the induction of COX-2 after ischemia, suggesting the involvement of NMDA receptors in stroke-induced COX-2 expression [134].

In an elegant study by the group of Dr. Costantino Iadecola [135], COX-2 mRNA was found to be upregulated in the ischemic hemisphere, but not contralaterally, beginning 6 h and persisting until 24 h after ischemia. They also found marked upregulation of COX-2 immunoreactivity at 12-24 h of reperfusion following 2 h of MCAO in the rat. Concentrations of $PGE_2$ were elevated in the stroke side by about 300% at 24 h [135]. Similar results were obtained in another investigation, where 12 h following 90 min of ischemia, COX-2 mRNA was induced in all layers in the neocortex, perifocal striatum, piriform cortex, dentate gyrus, and CA1-4 pyramidal cell layers in the hippocampus, as assessed using *in situ* hybridization [136]. The induction was also seen at the protein level using immunohistochemistry and immunoblotting. Most of the COX-2 positive cells were neurons. However, endothelial cells around the blood vessels in the infarcted core were strongly stained using the COX-2 antibody [136]. A similar increase in COX-2 in the vasculature was found in the piglet exposed to global ischemia or asphyxia, where marked immunofluorescence for COX-2 was observed in cerebral arteries and arterioles [137]. A more recent investigation found significant increase in COX-2 positive cells in the peri-infarct cortical area from 12 to 72 h after MCAO with reperfusion [138].

Using rat models of permanent and temporary MCAO, Kinouchi et al [139] found that COX-2 mRNA was not significantly induced in the ischemic core (lateral striatum) but strongly induced in the penumbral areas, comprising the medial striatum, periphery of MCA cortex as well as adjacent cortex (cingulated cortex). In brain areas distant from the ischemic territory, COX-2 mRNA was only induced bilaterally in the hippocampus. The substantial induction of COX-2 mRNA persisted in all regions even at 24 h after both permanent and transient MCAO [139].

In a more recent study, Yokota and colleagues [140] investigated the temporal and topographic profiles of COX-2 expression in permanent MCAO in rats. COX-2 mRNA increased significantly between 3 and 24 h of ischemia in the peri-infarct area compared to controls. In the ischemic core, significant increases in COX-2 mRNA levels were seen following 6 h of ischemia, which remained elevated up to 24 h. COX-2 immunoreactive neurons were found predominantly in the peri-infarct area, though elevations in the immunohistochemical staining of discrete neuronal populations were also observed in the ischemic core. Significant increases in $PGE_2$ levels were found in the ischemic hemisphere following 24 h of ischemia. The upregulation of COX-2 mRNA in the peri-infarct area persisted for at least 24 h after ischemia, as did the production of COX-2 protein, which led to significant increases in prostacyclin ($PGI_2$) as well as $PGE_2$ levels following 24 h of ischemia. In the ischemic core, increases in COX-2 mRNA persisted during the 24 h of ischemia, though significant increases in COX-2 protein were not found [140]. They observed that the magnitude of COX-2 activity and prostaglandin production was determined by the degree and duration of cerebral blood flow (CBF) reduction [140;141].

Another important mechanism of COX-2 induction following ischemia, as observed in rodents and non-human primates, is spreading depression [136;142-144]. Spreading depression is a wave of reduced spontaneous electrical activity that spreads across the brain. Repetitive spreading depression waves originate from the ischemic core and progress outward toward the penumbra, and are thus called ischemic depolarizations. This phenomenon is thought to be associated with the progressive neuronal damage observed during brain ischemia, and influences the expression of many genes including COX-2.

Unlike COX-2, COX-1 expression seems not to be affected by ischemic brain injury, as consistently reported by several groups using animal models of both global and focal cerebral ischemia [123;127;135;145;146]. However, Schwab and colleagues observed accumulation of COX-1-expressing microglial cells/macrophages in lesions of human focal ischemic stroke [85]. In support of this finding, a recent study has described significant increase in the number of cells positive for COX-1 in the peri-infarct frontoparietal cortex in a rat model of ischemic stroke [138]. Based on these results, it has been suggested that while the global levels of COX-1 may not increase following brain ischemia, it is possible that in those regions where there is proliferation and migration of microglia, there is indeed an increase in the expression of COX-1 [17;147].

As discussed earlier in this review, COX-2 expression is substantially increased following brain ischemia in rodents. However, these observations are not only restricted to experimental models. It is important to highlight that several reports have also described significant upregulation of COX-2 protein in post-mortem tissue obtained from humans with cerebral ischemia. In globally infarcted human brain, COX-2 protein was present in both neuronal and glial cells throughout the brain in accord with infarct topography and duration [148]. These authors suggested that the early induction of COX-2 may be involved in tissue injury, while the delayed induction in areas remote from the infarct may promote tissue scaring and remodeling [148]. In the acute stages of cerebral ischemic infarction, COX-2 protein was present in infiltrating neutrophils, vascular cells and neurons located at the border of the infarct, raising the possibility that COX-2 is involved in ischemic injury in the human brain [149]. In individuals who died two or more weeks after resuscitation (global brain ischemia), high expression of COX-2 protein was detected in macrophages and leukocytes at the center of the necrotic area. Reactive astrocytes and blood vessels were also COX-2 immunopositive [150]. In another study, Tomimoto and co-workers found strong immunoreactivity for COX-2 in microglia in the brains of patients who died from chronic cerebral ischemia such as Binswanger's disease [151]. Moreover, overexpression of COX-2 has also been shown in the ischemic neonatal human brain (neonates delivered after severe birth asphyxia) [152].

A wide body of evidence supports the notion that COX-2 induction could be responsible for a major component of tissue damage occurring in ischemia and represents and important target for treatment. However, it must be emphasized that COX-2 expression by itself does not lead to neuronal death since a variety of healthy neuronal populations throughout the CNS express COX-2 mRNA and protein under normal conditions [83;87] and COX-2 expression can be experimentally induced without causing death of these neurons [153].

In the study of Planas et al. [153], they found the same level of COX-2 induction in response to both mild ischemia (10 min of MCAO), which does not cause inflammation or cell death, and 1 h of stroke which leads to brain infarct. These data suggest that COX-2 would only mediate neuronal injury in the context of an inflammatory response. In support of this study, Yokota and colleagues found that in a primate model of permanent carotid artery embolization, COX-2 expression persisted even 24 h after ischemia in regions of the ipsilateral parietal cortex representing potentially viable tissue where cerebral glucose metabolic rate was preserved. This supports the notion that COX-2 induction is not always associated with cell death.

## ROLE OF CYCLOOXYGENASES IN BRAIN INJURY FOLLOWING ISCHEMIA

The relative contribution of each COX isoform to brain damage induced by ischemia has been extensively studied using pharmacological and genetic approaches. The availability of highly selective inhibitors of either COX-1 or COX-2 is a very powerful pharmacological tool to study the role that these specific isozymes play in the ischemic cascade. In addition, mice lacking each of these enzymes have been described, and extensively characterized in models of inflammation, cancer, pain, excitotoxicity, and cerebral ischemia.

### Pharmacological studies using COX inhibitors in models of brain ischemia

Numerous studies published several years ago showed that inhibition of COX using non-selective inhibitors protected against neuronal injury after ischemia in some animal models. Pre-treatment with indomethacin, an inhibitor of COX-1 and COX-2, has been shown to reduce infarct size and completely abolished the production of $PGD_2$ following focal brain ischemia [154;155]. It also diminished glutamate release from the ischemic cerebral cortex in a rat model of four-vessel occlusion, which results in global brain ischemia [156]. Indomethacin also ameliorated neuronal death in the hippocampal CA1 sector following 5 min of forebrain ischemia in gerbils [157;158], and reduced the number of TUNEL-positive CA1 pyramidal neurons in the same ischemia model [159-161]. However, using a similar dose, but a longer time of occlusion (10 min), indomethacin failed to reduce CA1 neuronal death [162]. This is in contrast to other reports in which edema formation is significantly reduced in the ischemic gerbil brain after treatment with indomethacin [163;164]. Piroxicam, a COX inhibitor widely used in the clinic, was also found to be very effective in reducing $PGE_2$ formation in the hippocampus and delayed neuronal death after global ischemia in both rats and gerbils [123;165]. Ischemic loss of CA1 neurons, assessed 7 days after common carotid artery occlusion in gerbils, is prevented by repeated treatment for 3 days with flurbiprofen [165].

Ibuprofen, another non-selective COX inhibitor, ameliorated delayed CA1 neuronal death in a model of global ischemia in rats [166;167], and reduced infarct size in temporary focal ischemia [168;169]. Interestingly, ibuprofen was not effective in reducing focal ischemic damage in a permanent MCAO model, even when administered at the same dose at which it was effective in transient MCAO [168]. Similarly, a very recent study has found that aspirin, given in repeated doses at a relatively high dose of 40 mg/kg, reduced infarct size in a transient [170;171], but not in a permanent, MCAO model in rats [171]. Of interest is the finding that aspirin reduces infarction and improves neurological recovery in temporary focal ischemia even when the first treatment is started up to 6 h after the onset of MCAO [172]. Moreover, KBT-3022, another non-selective COX inhibitor, was as effective as indomethacin in reducing CA1 neuronal cell death in a global ischemia model in gerbils [173]. In the same study, aspirin failed to reduce ischemic brain injury even at doses as high as 300 mg/kg [173]. Furthermore, the COX inhibitor tenoxicam significantly reduced hippocampal and striatal damage in rats subjected to bilateral common carotid artery occlusion for 45 min followed by 24 h of reperfusion [174]. It was also found that tenoxicam completely prevented ischemia-induced increase in myeloperoxidase activity (an index of infiltration of leukocytes) [174].

The COX-2 selective inhibitors NS-398 and valdecoxib and the non-selective inhibitor ketorolac have potent effects on cell death in cortical primary rat neuronal cultures subjected to hypoxia. These compounds reduced cell death by 70-95% at maximally effective doses [175].

It should be noted that treatment with non-selective COX inhibitors has not always resulted in neuroprotection. For example, indomethacin failed to increase the survival of CA1 hippocampal neurons in rats subjected to global ischemia [176], which is in contrast to what was found in gerbils [157]. Ibuprofen significantly increased the size of infarct in a model of permanent MCAO [168]. Furthermore, a very recent study showed that two COX inhibitors, indomethacin and NS-398, increased the proportion of animals displaying neuronal damage in a rat model of permanent cerebral hypoperfusion [177]. Findings from another report indicate that chronic inhibition of COX-2 with NS-398 only provided limited functional improvement following motor cortex stroke [178].

The chemical and pharmaceutical development of selective COX-2 inhibitors has fueled several investigations of their effects in models of cerebral ischemia. A wide body of evidence indicates that COX-2 inhibition confers neuroprotection in both global and focal ischemia. The protection is long-lasting and is also observed in rat and mouse models of permanent MCAO. More importantly, reduction in cell death correlated with improved recovery of the neurological function in ischemic animals treated with COX-2 inhibitors.

The COX-2 selective inhibitor NS-398 (20 mg/kg) reduced cortical, but not subcortical infarct volume at 3 days when administered starting 6 h following 2 h of MCAO in the rat. This was accompanied by a significant reduction in postischemic increase of $PGE_2$ production in the cerebral cortex [135]. Using the same rat model, Hara and co-workers found that the ischemic hemisphere had a significantly lower $PGE_2$ concentration compared with the contralateral hemisphere in the NS-398-treated group [179]. However, they observed no effect of NS-398 (10 mg/kg) on infarct size when rats were sacrificed after only 4 h of reperfusion. The discrepancy between these two reports [135;179] may be due to different time points at which the infarct volume was investigated. In addition, after 4 h of reperfusion, only the core of the lesion (mainly striatal areas) is overtly infarcted, and COX-2 does not seem to play an important role in the early death of the ischemic core [135], but rather is actively involved in the progression of the cortical injury in this rat model of focal ischemia [135;180;181]. In mice subjected to distal permanent MCAO, delayed administration of NS-398 (starting 18 h after the onset of stroke) was effective in diminishing infarct size and neurological deficits [182]. Surprisingly, there is a report indicating that NS-398 and nimesulide aggravate neuronal death by enhancing the increases in extracellular glutamate and intracellular $Ca^{2+}$ levels following oxygen-glucose deprivation in rat primary cortical neurons [183]. The mechanism(s) underlying this toxic effect of these COX-2 inhibitors is presently not known, and is in sharp contrast with data obtained in *in vivo* models of cerebral ischemia.

Nimesulide is a selective COX-2 inhibitor with the ability to rapidly and readily cross the blood-brain barrier [184;185]. It is structurally very similar to NS-398 and belongs to the class of acidic sulfonamides [44]. The neuroprotective efficacy of nimesulide has been extensively studied and overwhelming evidence indicates that it is a very promising neuroprotectant. Bilateral ligation of the common carotid arteries in the rat induces chronic cerebral hypoperfusion leading to white matter injury [186]. Treatment with nimesulide significantly attenuated white matter injury and microglial activation in this animal model [186]. Nimesulide was also very effective in reducing delayed neuronal death of hippocampal CA1 neurons following transient global ischemia in both gerbils [187] and mice [127]. Of interest is the finding that nimesulide rescued CA1 pyramidal neurons from ischemic death even when treatment was delayed until 24 h after ischemia, and the neuroprotective effect of nimesulide is still evident 30 days after the ischemic episode [187]. This suggests that COX-2 inhibitors confer long-lasting neuroprotection. Attenuation of oxidative stress seems to contribute to nimesulide's protective ability in global brain ischemia [188;189].

In focal ischemic brain injury, nimesulide significantly reduces infarct volume. It produced a more significant reduction in cortical than in striatal infarct after 1 h MCAO followed by 3 days of reperfusion [180]. Substantial protection by nimesulide was observed even when the first dose was given 24 hours after ischemic stroke, indicating a wide therapeutic window for neuroprotection with this COX-2 inhibitor [180]. Importantly, delayed treatment with nimesulide also improved neurological function as demonstrated by a better performance of treated rats in the rotarod test and lesser neurological deficit scores compared with vehicle-treated animals [180]. Stroke-induced elevation in $PGE_2$ levels in the cerebral cortex was completely prevented by nimesulide administration [180]. It has also been shown that nimesulide attenuates ischemic brain injury resulting from permanent MCAO in terms of reduction in infarct size and improvement of the neurological function [190]. In a very recent report, inhibition of COX-2 with nimesulide was found to effectively limit the blood-brain barrier breakdown, leukocyte infiltration, and vasogenic edema induced by focal cerebral ischemia [181]. Interestingly, in the same study, administration of the selective COX-1 inhibitor valeryl salicylate had no effect on the ischemic outcome, suggesting that COX-1 seems not to play an important role in the evolution of focal ischemic damage [181]. Using an embolic stroke model in rats, Wang et al. [191] confirmed the neuroprotective ability of nimesulide. It was found that administration of clinically relevant doses of nimesulide decreased neurological deficits, infarct volume, and brain edema. Hemorrhagic transformation was reduced by 64% with treatment of 12 mg/kg nimesulide. They also showed that the number of cells expressing matrix metalloproteinase (MMP)-9 and MMP-2 increased in the ischemic animals treated with vehicle, and significantly decreased in nimesulide-treated animals. This finding is of great relevance since MMPs have been shown to be actively involved in the detrimental pathological cascade resulting in ischemic injury, mainly due to their ability to disrupt most of the components of the blood-brain barrier [192-196]. The reduction in gelatinase content (MMP-2 and MMP-9) after ischemia in animals treated with nimesulide [191] helps to explain the protective effect of this COX-2 inhibitor against BBB breakdown and leukocyte infiltration, as shown in a previous study [181].

Advances in medicinal chemistry and considerable input from several pharmaceutical companies allowed the discovery of several highly selective COX-2 inhibitors. These agents have been tested in models of cerebral ischemia. In a seminal investigation by Nakayama and colleagues [123], it was convincingly demonstrated that COX-2 plays a role of paramount importance in the delayed neuronal death of CA1 hippocampal neurons following global ischemia. In this study, SC58125 was effective in preventing ischemia-induced elevation of hippocampal $PGE_2$ and delayed neuronal cell death [123]. Also in global ischemia, another highly selective COX-2 inhibitor, DFU, protected hippocampal neurons against transient forebrain ischemia even when administered several hours after the onset of reperfusion. DFU also reduced the increase in spontaneous locomotor activity seen in ischemic gerbils 7 days after 5 min of bilateral common carotid occlusion [197].

The relative contribution of each COX isoform to neurodegeneration and oxidative damage induced by global ischemia has been investigated using highly selective inhibitors of these enzymes [125]. It is of importance to emphasize that this study showed that either the selective inhibition of COX-2 with rofecoxib (Vioxx) or COX-1 inhibition with valeryl salicylate rescued CA1 hippocampal neurons from delayed ischemic death. It was also found in this investigation that inhibition of either COX-1 or COX-2 significantly limited oxidative damage to the hippocampus [125]. These data indicate that COX-1 seems also to be important in the neurodegenerative events triggered by global ischemia, and provided the first evidence indicating that both COX isoforms are involved in the progression of neuronal damage following global ischemia. These results are somehow unexpected since COX-1 is not induced after ischemia [123;135]. Nevertheless, the COX-1 inhibitor was as effective as the COX-2 inhibitor rofecoxib in attenuating delayed neuronal death after global ischemia [125]. This is in sharp contrast to findings from another study in which inhibition of COX-1 failed to modify brain injury induced by focal ischemia [181].

Pharmacological inhibition of COX-2 with SC58236 dose-dependently prevented $PGD_2$ increase and caused a significant reduction in the damaged area in a rat model of photothrombotic stroke [198]. More recently, Kelsen and co-workers found that administration of parecoxib, the water-soluble prodrug of valdecoxib (highly selective COX-2 inhibitor), produced a significant reduction in infarct volume when assessed 7 days after temporary MCAO in the spontaneously hypertensive rat [146]. Interestingly, they found a divided treatment response with responder and non-responder animals [146]. The reason(s) for this phenomenon was not clarified in the paper.

It should be noted that, at least in focal cerebral ischemia, the neuroprotective efficacy of COX-2 inhibitors is lost in mice lacking the inducible nitric oxide synthase (iNOS) gene [199]. This is a very interesting finding, and suggests that iNOS-derived nitric oxide production is required for COX-2 neurotoxicity in the context of cerebral ischemia [199]. In the focal ischemic stroke, COX-2 upregulation occurs with a time course very similar to that of iNOS [135;200]. Pharmacological inhibition or genetic ablation of iNOS significantly reduced postischemic increase in $PGE_2$ in the rat brain, indicating that nitric oxide formed by the action of iNOS drives COX-2 catalytic activity [201;202].

### Studies in COX-1 and COX-2 knockout mice subjected to brain ischemia

Investigations on COX-1 and COX-2 null mice have provided significant information on the roles that these isozymes play in ischemic brain injury. In line with the pharmacological studies, it was found that COX-2 knockout mice are less susceptible to MCAO, displaying a significant reduction in infarct size and $PGE_2$ accumulation in the ischemic cerebral cortex [145]. Similarly, COX-2-deficient mice showed a substantial reduction in hippocampal neuronal death after global ischemia in comparison with wild-type mice [127]. Conversely, neuronal overexpression of the human COX-2 gene in a transgenic mouse model exacerbates focal ischemic injury, as reported in an interesting study by Doré and colleagues [203]. Overexpression of COX-2 in transgenic mice resulted in a dramatic increase in infarct volume and post-ischemic production of $PGE_2$ and $PGF_{2\alpha}$ when compared with non-transgenic controls [203]. Furthermore, global ischemia resulted in significantly higher neuronal damage in the CA1 region of the hippocampus of transgenic mice with neuronal overexpression of COX-2 than in wild-type controls [204], indicating a deleterious role of COX-2 in ischemic neuronal damage. In addition, primary cortical neurons prepared from COX-2 deficient mice are more resistant to hypoxia than control cells [175].

On the other hand, genetic disruption of COX-1 decreases resting cerebral blood flow [205] and results in enhancement of infarct size and neurological deficits following distal MCAO [206]. However, these detrimental effects of COX-1 knockout were not confirmed in a model of permanent endovascular occlusion of the middle cerebral artery, where no difference on stroke outcome was observed between COX-1 null mice and wild-type controls [207]. In support of this, another report showed no modification in infarct size between COX-1 knockout and wild-type mice in a photothrombotic model of stroke [208]. Moreover, selective inhibition of COX-1 was proven not to modify infarct size or neurological deficits in rats subjected to transient ischemic stroke [181]. On the contrary, Lin et al. [209] found a neuroprotective effect against focal ischemic injury after infusion into the rat lateral ventricle of recombinant adenoviruses containing either COX-1 or COX-1 plus prostacyclin synthase (PGIS). These apparently contradictory studies suggest that more work should be done in order to clarify the role of COX-1 in cerebral ischemia.

## CYCLOOXYGENASE INHIBITION IN MODELS OF EXCITOTOXIC BRAIN INJURY

Excitotoxicity and the resulting alteration of intraneuronal $Ca^{2+}$ homeostasis are among the most important pathophysiological events following brain ischemia [11]. These early processes propagate the cascade of harmful events leading to cerebral infarction [12;13]. As mentioned before, COX-2 expression is increased after activation of ionotropic glutamate receptors, and it has been proposed that enhanced COX-2 activity mediates, at least in part, the neurotoxicity associated to over-excitation of glutamate receptors after ischemia. However, there are conflicting data in this regard, as described below.

It has been demonstrated that COX-2 deficient mice displayed a reduced damage following intracerebral injection of NMDA [145;210], and administration of selective COX-2 inhibitors significantly attenuated this type of injury [145;210;211]. It has also been shown that the non-selective COX inhibitor, naproxen, significantly reduced brain edema and neuronal loss in a mouse model of intrahippocampal injection of NMDA [212]. Neuroprotective efficacy was still evident when naproxen administration was started 6 h after NMDA microinjection [212]. Furthermore, COX-2 selective blockade, but not COX-1 inhibition, has been demonstrated to effectively reduce brain damage caused by direct injection of the excitotoxins quinolinic acid [213] and quisqualic acid [214;215].

Using mixed mouse cortical cultures containing both neurons and astrocytes, Hewett and co-workers [116] found that flurbiprofen, a nonselective COX-1/COX-2 inhibitor, and the COX-2 selective inhibitor NS-398 prevented NMDA-induced production of prostaglandins and attenuated neuronal death. COX-1 inhibition with valeryl salicylate had no effect on neuronal death. This study also showed that the newly synthesized COX-2 protein is the one that contributes to NMDA neurotoxicity. This conclusion was based on two experimental findings: 1) inhibition of total COX activity with aspirin before NMDA exposure failed to prevent cell death; and 2) treatment with flurbiprofen after NMDA significantly attenuated neuronal death [116].

Two structurally distinct selective COX-2 inhibitors (NS-398 and APHS) were found to confer neuroprotection against NMDA-mediated excitotoxicity in an *in vitro* neuronal culture. The neuroprotective effect of APHS was reversed by the administration of $PGE_2$ or by a concentration-dependent treatment with 17-phenyl-trinor-$PGE_2$, which is a relatively specific agonist of EP1 and EP3 prostanoid receptors [216].

Interestingly, COX inhibitors do not protect against the neurotoxicity induced by the excitotoxin kainate, suggesting that the protection conferred by COX-2 inhibition is specific for NMDA [116]. In support of these data, another study performed in cerebellar granule neurons demonstrated that the highly selective COX-2 inhibitor, DFU (5,5-dimethyl-3-(3-fluorophenyl)-4-(4-methyl-sulphonyl) phenyl-2(5H)-furanone, completely protected cultured neurons against glutamate-mediated toxicity, and resulted in a significant attenuation of apoptotic cell death induced by an excitotoxic concentration of NMDA. In contrast, no protection from kainate-mediated neurotoxicity was observed [117]. Nevertheless, there is a report indicating that in primary cortical neurons, both indomethacin (non-selective inhibitor) and aspirin (COX-1 preferential inhibitor) reduced kainate-induced neuronal cell death, while NS-398 (COX-2 selective) had no effect. However, in primary hippocampal neurons, COX-2 inhibition prevented both kainate-induced cell death and $PGE_2$ synthesis [217].

Administration of COX inhibitors to animals injected systemically with kainate has produced mixed results. Pre-treatment of mice with the COX-2 selective inhibitor NS-398 increases hippocampal neuronal cell death and enhances mortality in the kainic acid model [218]. A very recent study also showed that COX-2-deficient mice, but not COX-1 knockout mice, exhibit an increased sensitivity to kainate-induced seizures and are more susceptible than COX-2 wild-type animals to excitotoxic neuronal damage [219]. They also found that chronic administration of celecoxib (COX-2 selective inhibitor) recapitulated the increased susceptibility to kainate neurotoxicity observed in COX-2-null mice [219]. These findings are not in line with data from another investigation demonstrating that transgenic mice with neuronal overexpression of human COX-2 showed a dramatic increase in neuronal damage following kainate injection, and glutamate excitotoxicity is potentiated in primary cortico-hippocampal neurons derived from these transgenic mice [220].

These apparent discrepancies in data from *in vivo* studies using the kainate model might be the result of several factors, including the timing of administration of the COX-2 inhibitor. Treatment with COX-2 inhibitors before kainate aggravates neuronal death and seizure severity [218;221], but post-treatment with COX-2 inhibitors corrected learning and memory deficits and prevented excitotoxic neuronal death in the hippocampus [222-225]. These findings suggest that inhibition of newly synthesized COX-2 induced by excitotoxins may confer protection, but blockade of basal COX-2 activity may be detrimental following kainate excitotoxicity.

## MECHANISMS OF TOXICITY OF ENHANCED COX ACTIVITY FOLLOWING CEREBRAL ISCHEMIA

There are several factors responsible for the cytotoxicity of COX-2 in the setting of cerebral ischemia, which result in neuronal injury. The increased production of free radicals and $PGE_2$ are among the most recognized mechanisms of toxicity linked to increased COX-2 activity. However, there are also two other processes, modulated by COX-2, which could potentially lead to neuronal death: 1) promotion of cell cycle activity by increasing cyclin D1 expression, and 2) metabolism of endocannabinoids (**Figure 2**).

Cerebral ischemia results in a substantial increase in the availability of arachidonic acid, the substrate for the COX enzymatic pathway. A wide body of experimental evidence supports the theory that COX catalytic activity in linked to the production of free radicals. Oxidative stress is considered to be one of the major determinants of ischemic neuronal death [226-228]. Detailed biochemical investigations have demonstrated that free radicals are indeed produced by the peroxidase step of the COX reaction in which $PGG_2$ is converted to $PGH_2$ [23;229-234]. Although it has become customary to consider reactive oxygen species (ROS), and specifically superoxide anion ($O_2\cdot-$), to be the primary radical produced by COX activity during inflammation, there is no direct evidence for this notion. On the contrary, the two major types of radicals so far known to be involved in COX activity are tyrosyl radicals on proteins [23] and carbon-centered radicals on the substrate arachidonic acid [232]. Due to the characteristically short half-life of free radicals, and the technical difficulties associated with their direct measurement in biological systems, there is still debate on the chemical nature of the free radical(s) involved in COX-2-mediated oxidative stress during inflammation and cerebral ischemia. However, there is an overwhelming line of evidence indicating that enhanced COX activity following cerebral ischemia and excitotoxicity is associated with oxidative damage.

Pharmacological inhibition of COX-2 with either nimesulide or rofecoxib resulted in a significant reduction in measures of oxidative injury in the hippocampus following global cerebral ischemia in gerbils. These COX-2 inhibitors prevented ischemia-induced glutathione depletion and the increase in lipid peroxidation, as assessed by the levels of lipid hydroperoxides, malondialdehyde (MDA) and 4-hydroxy-alkenals [125;188]. In a rat model of global forebrain ischemia, nimesulide treatment reduced lipid peroxidation and prevented the depletion of reduced glutathione following reperfusion [189]. Similarly, indomethacin treatment significantly reduced 8-hydroxy-deoxyguanosine (8-OH-dG), a highly sensitive marker of DNA oxidation, in the ischemic hippocampus [161]. In *in vivo* models of excitotoxicity, COX-2 has been also demonstrated to be a significant source of free radicals. Treatment with nimesulide significantly reduced oxidative injury in the rat hippocampus after systemic kainate injection [235]. A microdialysis study in the hippocampus of freely moving rats showed that the COX inhibitors flurbiprofen and indomethacin, or the COX-2 selective inhibitor NS-398 effectively reduced 8-epi-$PGF_{2\alpha}$ (15-$F_{2t}$-IsoP), a reliable marker of free radical-mediated lipid peroxidation [236], following infusion of NMDA [237]. Interestingly, using the same rat model of excitotoxicity, inhibition of COX-1 with SC-560 significantly attenuated the increase in hippocampal 8-epi-$PGF_{2\alpha}$ levels induced by NMDA [238]. In support of this study, it has been reported that the COX-1 inhibitor valeryl salicylate reduced measures of oxidative stress in the gerbil hippocampus following temporary global ischemia [125]. This suggests that COX-1 may also contribute to oxidative injury following excitotoxicity and brain ischemia.

Data from *in vitro* studies indicate that enhanced COX activity results in oxidative stress in neuronal and microglial cells exposed to inflammatory conditions. Inhibition of COX-2 activity with SC-58125, COX-1 inhibition with aspirin, or blockade of COX-1 and COX-2 with indomethacin significantly reduced the formation of 8-epi-PGF$_{2\alpha}$, a biomarker for lipid peroxidation, in human neuronal cells exposed to IL-1$\beta$ [239]. Furthermore, treatment with indomethacin prevented glutathione depletion in a rat model of neuroinflammation induced by intracerebral injection of TNF-$\alpha$ [240].

Microglial activation is actively involved in brain ischemia and neurodegeneration [55;241-245]. In LPS-activated microglia, selective inhibition of either COX-2 or COX-1 activity significantly attenuated the formation of free radicals and the ensuing lipid peroxidation [53;246]. In mouse microglial cells subjected to anoxia/reoxygenation, inhibition of COX with indomethacin attenuated the production of free radicals and diminished the oxidative damage to lipids and proteins [247]. Interestingly, the highly COX-1 selective inhibitors SC-560 and valeryl salicylate completely prevented PGE$_2$ production and lipid peroxidation in LPS-activated microglial cells [246;248], despite the fact that COX-1 is not induced by LPS in these cells. Also *in vivo*, COX-1 inhibition attenuated LPS-induced oxidative stress and brain injury [249]. These recent findings suggest that the contribution of COX-1 to neurotoxicity following ischemic and inflammatory injury should not be ignored.

Results from all the above-mentioned investigations convincingly indicate that COX activity is a significant source of free radicals during cerebral ischemia and excitotoxicity. However, a couple of reports from Dr. Iadecola's group have challenged this idea. They found that COX-2 selective inhibition with NS-398 or COX-2 genetic ablation diminished lesion volume, but did not reduce ROS formation in mice with NMDA-induced neocortical lesions. The protection conferred by NS-398 is counteracted by a stable analog of PGE$_2$ [210]. Authors concluded that COX-2-derived prostanoids, rather than ROS, mediate the COX-2-dependent component of the damage mediated by the activation of NMDA receptors. In a similar study, but using a mouse MCAO model of focal ischemia, they showed that NS-398 attenuated the production of PGE$_2$ and reduced infarct size, but did not affect ROS production. Similarly, ROS formation was not reduced in COX-2 knockout mice [250]. Data from both of these studies should be carefully interpreted. They used hydroethidine fluoromicrography to assess ROS formation. This method only detects the production of ROS, mainly O$_2$·$^-$ [251-253]. Since emerging evidence indicate that O$_2$·$^-$ is not the major radical formed by COX activity [23;232], the apparent discrepancies between these two reports [210;250] and results from other studies [125;161;188;237;238;249] could be explained by the different methodological approach utilized. Unlike Iadecola's studies, other investigations have measured markers of oxidative stress (e.g., glutathione levels, MDA, lipid hydroperoxides, isoprostanes, protein and DNA oxidation) instead of direct ROS detection. As mentioned before, the identity of the radical species formed during COX catalysis has not been established. Irrespective of the chemical nature of the radical formed by COX, by measuring biomarkers of oxidative damage, several studies have convincingly shown that increased COX activity contributes to oxidative stress following ischemic and excitotoxic brain injury [125;161;188;237;238]. It has also been demonstrated that oxidative stress directly induced neuronal COX-2 transcription [254-256]. This could lead to a vicious cycle associated with neurotoxicity.

Production of pro-inflammatory prostanoids, and more specifically PGE$_2$, is another pivotal mechanism of neurotoxicity associated to increased COX activity after ischemia and excitotoxicity. Work by Manabe and co-workers indicated that PGE$_2$, but not PGF$_{2\alpha}$, mediates, at least in part, the neurotoxicity of COX-2 following microinjection of NMDA into the neocortex in mice [210]. In addition, recent data indicate that the activation of the PGE$_2$ receptor EP1 underlie the neurotoxicity of COX-2. Pharmacological blockade or gene inactivation of EP1 receptors reduce neuronal damage induced by focal cerebral ischemia, excitotoxicity, and oxygen-glucose deprivation [257;258]. More recently, it has been shown that activation of the EP3 receptor exacerbates ischemic and excitotoxic brain injury *in vivo* [259].

It has long been described that PGE$_2$ may mediate both detrimental and beneficial effects on neuronal survival. Addition of PGE$_2$ induced apoptotic cell death in rat cortical and hippocampal neurons [260-263], and PGE$_2$ has been shown to increase glutamate release from astrocytes, which may potentiate excitotoxicity in the setting of cerebral ischemia [264]. However, it should be emphasized that numerous reports describe neuroprotective effects of $PGE_2$ against excitotoxicity and cerebral ischemia [265-267]. The type of $PGE_2$ receptor activated seems to determine whether $PGE_2$ exerts neuroprotection or neurotoxicity. Activation of either EP2 or EP4 results in neuroprotection against NMDA-mediated excitotoxicity [267-269], ischemia *in vivo* [267;270], and oxygen-glucose deprivation *in vitro* [267]. Binding of $PGE_2$ to EP1 or EP3 results in neurotoxicity in the setting of excitotoxic or ischemic events, as emerging evidence indicates [257-259;271].

The notion that $PGE_2$ is an important protagonist in the neuroinflammatory cascade which follows cerebral ischemia is further supported by a very recent investigation demonstrating that mice deficient in microsomal $PGE_2$ synthase-1 (mPGES-1), the most important isoform responsible for $PGE_2$ synthesis during neuroinflammation [272;273], displayed a significant reduction in ischemic brain injury as compared to wild-type animals [274]. In mPGES-1 null mice, in which the cortical postischemic production of $PGE_2$ was completely absent, the infarction, edema, apoptotic cell death, and caspase-3 activation after ischemia were all reduced compared with those in wild-type mice [274]. Of great interest was the finding that the improved outcome observed in mPGES-1 knockouts after focal ischemia were reversed to almost the same severity as wild-type animals by intracerebroventricular microinjection of $PGE_2$ into knockout mice [274].

Another mechanism by which COX-2 mediates the neurotoxic effect in ischemic brain injury is through promotion of the cell cycle activity [275;276]. Post-mitotic neurons are devoid of the capacity to divide or replicate, and are believed to be in the state of extended G0 phase. There are several stimuli which promote these post-mitotic neurons to divide. However, the lack of complete cell cycle machinery is lethal and such neurons undergo apoptotic cell death [275;277]. G1 cyclins are involved in promoting cell cycle progression in the G1 stage in dividing cells. Cyclin D1 forms molecular complexes with either cyclin-dependent kinase-4 (CDK4) or CDK6. These complexes lead to phosphorylation of the retinoblastoma tumor suppressor protein (Rb), a nuclear protein. Phosphorylation of Rb results in the release of the transcription factor E2F, which commits the cell to proliferation. It has been found that inhibition of cyclin D1/CDK4 activity protects neurons from apoptosis, and up-regulation of cyclin D1 induces neuronal cell death [278-283]. Cultured neurons subjected to hypoxia showed an increase in the expression of cyclin D1. COX-2 inhibition with NS-398 significantly reduced hypoxia-induced cyclin D1 protein expression. These findings were also replicated in the *in vivo* situation, where SC-58125, a highly selective COX-2 inhibitor, significantly attenuated the increase in cyclin D1 expression after focal cerebral ischemia in rats [275]. In cortico-hippocampal primary neuronal cultures, the phosphorylation of Rb is induced in neurons overexpressing human COX-2 during response to glutamate-mediated excitotoxic apoptotic damage [276;284]. Together, these results indicate that COX-2 activity may induce neuronal cell death following ischemia by inducing the expression of cyclin D1 and unsuccessful entry into the cell cycle.

The active principles of *Cannabis sativa L*, known popularly as marijuana, are a group of terpenophenolic compounds termed cannabinoids, which bind to the receptors CB1 and CB2. The endogenous ligands for these receptors are called endocannabinoids, and are physiologically produced by several cell types in the body. Relative recent investigations have shown that COX-2, but not COX-1, is able to oxygenate the endocannabinoids 2-arachidonoyl glycerol (2-AG) and arachidonoyl ethanolamide (anandamide or AEA) to yield new types of prostaglandins: prostaglandin glyceryl esters and prostaglandin ethanolamides [64;65;285;286]. The endocannabinoid signaling through CB1 and/or CB2 receptors has been reported to confer protection against cerebral ischemia [287-292]. Thus, by decreasing the levels of endocannabinoids, increased COX-2 activity may block this endogenous protective response of the brain to ischemia. Furthermore, a very recent evidence indicates that $PGE_2$ glyceryl ester, a major COX-2 oxidative metabolite of 2-arachidonoyl glycerol, potentiated excitatory glutamatergic synaptic transmission and induced apoptotic cell death when added to primary hippocampal neurons [285]. However, it remains to be determined whether these mechanisms contribute to the neurotoxic effect of COX-2 in the scenario of brain ischemia/reperfusion.

## CRITICAL APPRAISAL OF THE THERAPEUTIC POTENTIAL OF COX-2 INHIBITORS IN CEREBRAL ISCHEMIA

As described earlier in this article, neuronal injury following ischemia is attenuated by COX-2 inhibitors in several animal models, indicating that these agents may be of therapeutic benefit in the management of patients with cerebral ischemia. Relevant from the therapeutic point of view, numerous investigations demonstrate neuroprotection even when the COX-2 inhibitor is administered in a delayed fashion. This is of great importance because most patients arrive in the emergency room several hours after the onset of ischemia, a time at which most available therapeutic options are no longer effective. In addition, neuroprotection exerted by COX-2 inhibitors is long-lasting and results in improvement of the neurological function.

Despite all these promising lines of experimental evidences, clinical trials evaluating efficacy of COX-2 inhibitors in patients with stroke are very unlikely. This is mainly due to recent concerns about the increased cardiovascular risk after chronic treatment with this class of pharmacological agents [74;293-298]. It has now been suggested that all COX inhibitors could potentially result in increased risk of cardiovascular events [295;299-301]. These drugs include the widely used ibuprofen, naproxen, diclofenac, celecoxib, among others. A recent clinical study indicates that COX-2 selective inhibitors may be differ in their potential to cause harmful cerebrovascular effects. Use of rofecoxib and etoricoxib, but not of celecoxib was associated with an increased risk of ischemic stroke. Odds ratios of ischemic stroke appeared to increase with higher daily dose and longer duration of rofecoxib and etoricoxib use. Authors conclude that for the risk of ischemic stroke additional pharmacological properties of individual COX-2 inhibitors may be important [302]. It is noteworthy to mention that the increased toxicity of COX-2 inhibitors is observed after prolonged administration in patients at risk to develop cardiovascular events [303-306] and the relative cardiovascular risks of NSAIDs are similar in magnitude to other currently prescribed therapies [307].

Thus, the possible short-term benefits of COX-2 inhibition in patients with an acute ischemic stroke should not be overlooked. Based on the pre-clinical data, these agents could result in neuroprotection and significant improvement of the neurological function when given therapeutically for just few hours/days after ischemia in selected patients. As with any drug, the risk/benefit ratio should be always considered and carefully evaluated by the physician.

Very little information is available on the long-term effects of COX inhibition in the ischemic brain. This is an important limitation of most of the experimental studies evaluating the efficacy of COX inhibitors in models of brain ischemia. Blockade of COX activity may alter the neurogenic response, and this could negatively impact the endogenous repairing mechanisms which follow cerebral ischemia. It has been demonstrated that repeated treatment with aspirin significantly reduced ischemia-induced proliferation of cells in the dentate gyrus of adult gerbils after global brain ischemia [308]. In a similar study, Sasaki and colleagues [309] found that both indomethacin and the COX-2 inhibitor NS-398 significantly blunted the enhancement of dentate gyrus proliferation of neural progenitor cells following ischemia in the mouse. Furthermore, in the post-ischemic dentate gyrus of heterozygous and homozygous COX-2 deficient mice, proliferating bromodeoxyuridine-positive cells were significantly fewer than in wild-type littermates [309]. These results demonstrate that COX-2 is an important modulator in enhancement of proliferation of neural progenitor cells after ischemia.

Additionally, very recent pieces of evidence suggest that COX-2 plays an essential role in cerebral ischemic preconditioning, and could potentially have protective effects [310;311]. The beneficial roles of COX-2-derived metabolites in neurogenesis and ischemic tolerance should be carefully considered when developing neuroprotective strategies based on COX-2 inhibition or blockade of prostanoid receptors.

**CONCLUSIONS AND FUTURE DIRECTIONS: TARGETS DOWNSTREAM OF COX INHIBITION**

Due to emerging evidence suggesting that inhibition of COX-2 may result in increased risk of cardiovascular events, efforts have been made to elucidate the downstream effectors of COX-2 neurotoxicity with the hope of developing new therapeutic agents lacking the detrimental side effects. Modulation of $PGE_2$ effects through EP receptors appears to be a promising target.

Pioneer work from the labs of Drs. Iadecola and Dore indicate that pharmacological blockade of the $PGE_2$ receptor EP1 using SC51089 or EP1 genetic deletion results in substantial neuroprotection against excitotoxic and ischemic cerebral damage [257;258;271]. Work by Kawano et al. [257] clearly showed that blockade of EP1 receptors reduced brain injury when administered 6 hours after MCAO, suggesting that EP1 receptor inhibition may be a viable therapeutic option in ischemic stroke [257]. It has also been demonstrated that agonists of EP2 or EP4 receptors exert neuroprotective effects in experimental models of excitotoxicity and cerebral ischemia [267-270].

Although manipulation of $PGE_2$ signaling using an agonist/antagonist approach seems to be a valid therapeutic strategy, none of these pharmacological agents are clinically available, and thorough pharmacodynamic and toxicological studies remain to be performed before patients suffering from stroke will benefit from the therapeutic use of these compounds.

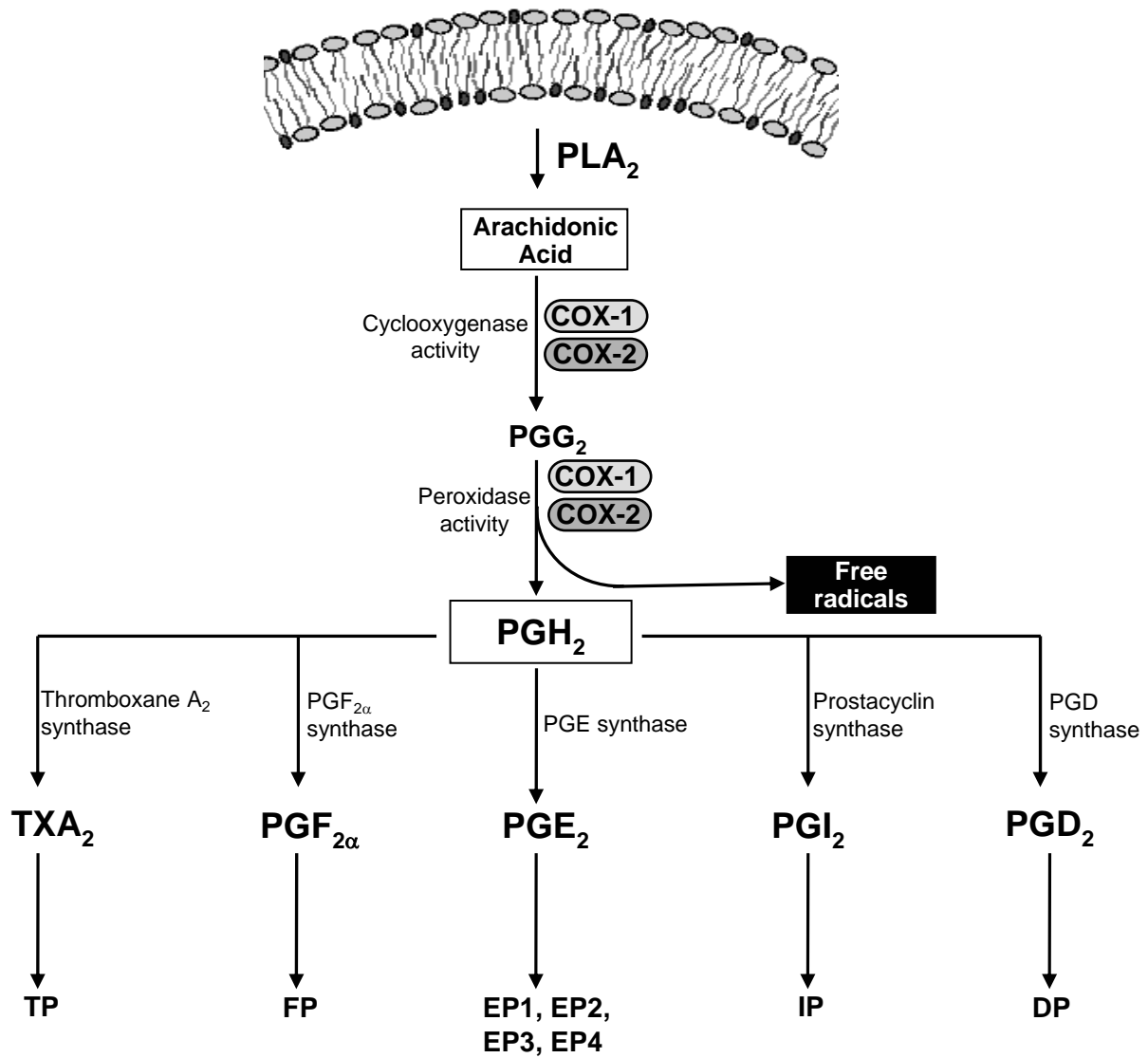

**Fig. 1.** Arachidonic acid metabolism through COX-1 or COX-2 results in the production of different prostanoids, which exert their biological effects via specific receptors, coupled to various signaling transduction pathways. It should be noted that the peroxidase reaction of COX is linked to the production of free radicals. For more details, please refer to text.

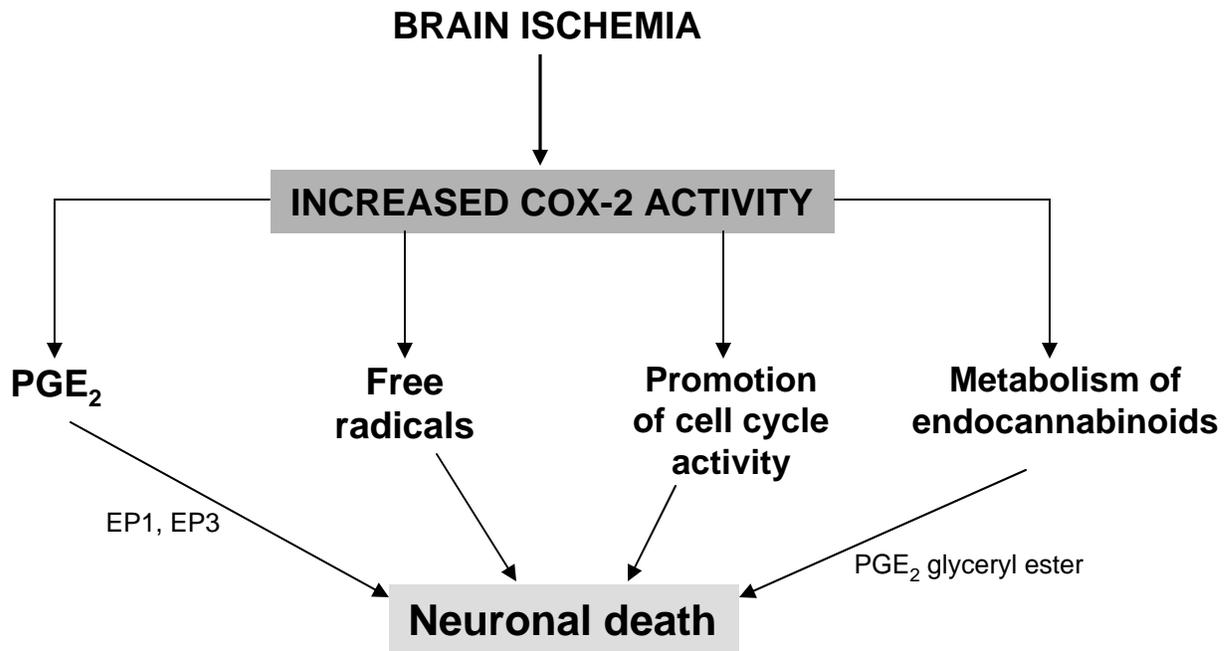

**Fig. 2.** Proposed mechanisms associated with the neurotoxicity of increased COX-2 activity following brain ischemia.